# Realization of ultrabroadband THz/IR photoresponse in a bias-tunable ratchet photodetector

Peng Bai[1, 6], Xiaohong Li[2, 6], Ning Yang[1], Weidong Chu[1, *]

Xueqi Bai[2], Siheng Huang[2], Yueheng Zhang[2, *], Wenzhong Shen[2]

Zhanglong Fu[3], Dixiang Shao[3], Zhiyong Tan[3], Hua Li[3], Juncheng Cao[3]

Lianhe Li[4], Edmund Harold Linfield[4]

Yan Xie[5], Ziran Zhao[5, *]

1. *Institute of Applied Physics and Computational Mathematics, Beijing 100088, China*
2. *Key Laboratory of Artificial Structures and Quantum Control, School of Physics and Astronomy, Shanghai Jiao Tong University, Shanghai 200240, China*
3. *Key Laboratory of Terahertz Solid-State Technology, Shanghai Institute of Microsystem and Information Technology, Chinese Academy of Sciences, Shanghai 200050, China*
4. *School of Electronic and Electrical Engineering, University of Leeds, Leeds LS2 9JT, UK*
5. *Department of Engineering Physics, Tsinghua University, Beijing 100084, China*
6. *These authors contributed equally: Peng Bai, Xiaohong Li*

*Corresponding authors, E-mail addresses: chu_weidong@iapcm.ac.cn, yuehzhang@sjtu.edu.cn, zhaozr@mail.tsinghua.edu.cn

## Abstract

High performance Terahertz (THz) photodetector has drawn wide attention and got great improvement due to its significant application in biomedical, astrophysics, nondestructive inspection, 6th generation communication system as well as national security application. Here we demonstrate a novel broadband photon-type THz/infrared (IR) photodetector based on the GaAs/Al$_x$Ga$_{1-x}$As ratchet structure. This kind of photodetector realizes a THz photon-response based on the electrically pumped hot hole injection and overcomes the internal workfunction related spectral response limit. An ultrabroadband photoresponse from 4 THz to 300 THz and a peak responsivity of 50.3 mA/W are realized at negative bias voltage of -1 V. The photodetector also presents a bias-tunable photon-response characteristic due to the asymmetric structure. The ratchet structure also induces an evident photocurrent even at zero bias voltage, which indicates the detector can be regard as a broadband photovoltaic-like detector. The rectification characteristic and high temperature operation possibility of the photodetector are also discussed. This work not only demonstrates a novel ultrabroadband THz/IR photodetector, but also provides a new method to study the light-responsive ratchet.

# 1. Introduction

Different kinds of Terahertz (THz) photodetectors have caught more attention and been intensively explored recently thanks to the diverse range of applications in biomedical, astrophysics, national security and next generation communication system (6G) [1-5]. Although great developments have been made during the last several decades, THz photodetectors still suffer from different drawbacks limiting their practical applications. The thermal photodetectors such as bolometer, Golay cells, and pyroelectric photodetector are highly developed and now all commercially available [6]. But their sensitivity and response time are often incompatible. The modulation frequency is usually < 100 Hz for nearly all thermal photodetectors even though the real-time imaging has already been achieved using micro-thermal photodetector array [7]. The Schottky-barrier diode THz direct photodetectors could realize sub-microsecond response time. However, the limited response frequency range (< 2 THz) is the main drawback to their wider applications [8]. Further extension of the frequency coverage and array integration of Schottky diodes would be highly desired. Field effect transistor (FET)-based THz photodetector is totally compatible to standard silicon microelectronics processing technology and may be a promising candidate for large scale imaging arrays [7, 9]. The narrow response bandwidth is still the biggest obstacles for wide application of Tera-FET.

Photon-type photodetectors exhibit adjustable response range, good signal-to-noise performance and very fast response. THz quantum well photodetectors (THz-QWPs) have been proved an excellent photon-type photodetector benefiting from the fast response speed, high sensitivity, and flexible and adjustable photon-response range [10-12]. However, due to the limitation of the selection rule of inter-subband transition (ISBT), the n-type THz-QWP requires optical coupling structure design [13-15]. Moreover, the narrow band detection characteristic makes QWPs optical filter free in some applications, but limits their spectrum coverage on the other side. The homojunction or heterojunction interface workfunction internal photoemission (HIWIP or HEIWIP) photodetectors have been a competitive THz photodetector because of their normal incidence response mechanism, wide spectrum response coverage, and tailorable cutoff frequency [16, 17]. Unfortunately, the low activation energy (~10 meV) of the IWIP results in large dark current and requires extremely low temperature (~4 K) operation condition [18]. Another alternative is quantum dot THz photodetector, which could realize normal-incidence response and high temperature operation. Nevertheless, the reliability and repeatability of the quantum dot material growth is still a great challenge [19]. The optical pumped hot-hole effect photodetector (OPHED) is based on the hot-cold hole energy transfer mechanism that overcomes the bandgap spectral limit and realizes THz detection [20]. This type of detection mechanism enables a designable detection wavelength by adjusting the potential barrier while simultaneously suppressing the dark current and noise [21]. However, the external optical excitation dependent hot holes injection is the prerequisite for THz detection, which greatly increases the complexity of OPHED. Moreover, the reported work displayed a responsivity of only a few μA/W that indicates the need for further optimization [20].

Ratchets are nonequilibrium systems that break spatial inversion symmetry which could be realized in variety systems [22]. Producing directional motion of micro-particles by using a time-varying input of nondirectional or random perturbations such as radiation, heat or applied AC fields without using a bias is the significant feature of the ratchet effect [23]. This concept originates from the cognitive process of symmetry laws and the second law of thermodynamics. Extensive research has been carried out in the fields of photovoltaics and molecular motors since it came into the 21st century [24-26]. Radiation detection (sensing) is one of the main suggested applications where incident radiation is rectified in the ratchets to produce currents [27]. Recently, the THz detection has been realized in the electron mobility transistor based ratchet, graphene sheet based ratchet and silicon-based ratchet respectively, which indicates the ratchet effect is a potential solution for THz radiation detection even though there is no reported direct experimental broadband THz photon-response study [28-30]. The ability to work under zero bias and the bias-adjustable photoresponse range make the ratchet terahertz photodetectors have great potential in the field of high-temperature and broadband detection. A potential concept based on GaAs/ $Al_xGa_{1-x}As$ quantum ratchet was demonstrated by C. Phillips in 2012[31] and realized by Anthony *et.al* in 2018, in which the ratchet band served as an intermediate state to absorb long wavelength photons and contribute to the photocurrent, thereby increasing the efficiency of solar cells [32]. The mature material growth technology and large bandgap difference between GaAs and AlAs give us a great freedom to construct unique type ratchet-like potential by band engineering to realize THz photon-response [33].

In this work, we propose a novel photon-type THz photodetector based on the GaAs/ $Al_xGa_{1-x}As$ ratchet structure. The photodetector allows normal incidence excitation thus bypassing the need for a grating coupler required for ISBT-based THz detectors. An ultrabroadband photoresponse from 4 THz to 300 THz covers a much wider range compared with the other GaAs-based photon type photodetectors. The THz response based on the electrically pumped hot hole injection breaks the spectral response limit. Moreover, an evident photocurrent is observed even at zero bias voltage and the response spectra exhibit strong bias dependent characteristics. The THz/IR response mechanism and photodetector performance are studied systematically. The peak responsivity of the proposed photodetector increases by three orders of magnitude in contrast to the OPHED. Additionally, the ratchet photodetector exhibits pronounced rectification behavior at the temperature below 77 K owing to the ratchet effect. The dark current of the detector is also much lower than the other photon-type THz photodetectors due to the high inherent barrier design, which makes it possible to improve the operation temperature.

## 2. Results

### 2.1 Device structure

The schematic structure of the ratchet photodetector device is displayed in Fig.1(a) (see the Method Section and Supplementary Information for the wafer details). The band diagram of ratchet photodetector under zero bias is shown in Fig.1(b). The basic detection cell is presented in the dashed box which is a four-layer structure consisting of injector layer, graded barrier layer, absorber layer and constant barrier layer. The barrier shape also represents the aluminum component (x) of the $Al_xGa_{1-x}As$. The bottom contact layer also served as the injector layer of the first period of the ratchet photodetector, which causes the first injector thicker than the other periods. The internal workfunction ($\Delta_B$) of the barriers in the detector structure are calculated as 67 meV and 120 meV respectively according to the high density (HD) theory (see Supplementary Information for the workfunction calculation of the GaAs/AlGaAs heterojunction) [16, 34]. The photoluminescence (PL) spectrum (as is shown in Fig.1(c)) of the quantum ratchet photodetector at room temperature presents broadband radiative recombination PL peaks. The multi-peak characteristic is analyzed using Lorentz fitting. The fitted peaks at 875 nm and 904 nm correspond to the interband transition and conduction-band to impurity-band transition of Be doped GaAs. The fitted multiple $Al_xGa_{1-x}As$ peaks indicate the gradual change of aluminum composition of the graded $Al_xGa_{1-x}As$ barrier in the quantum ratchet structure.

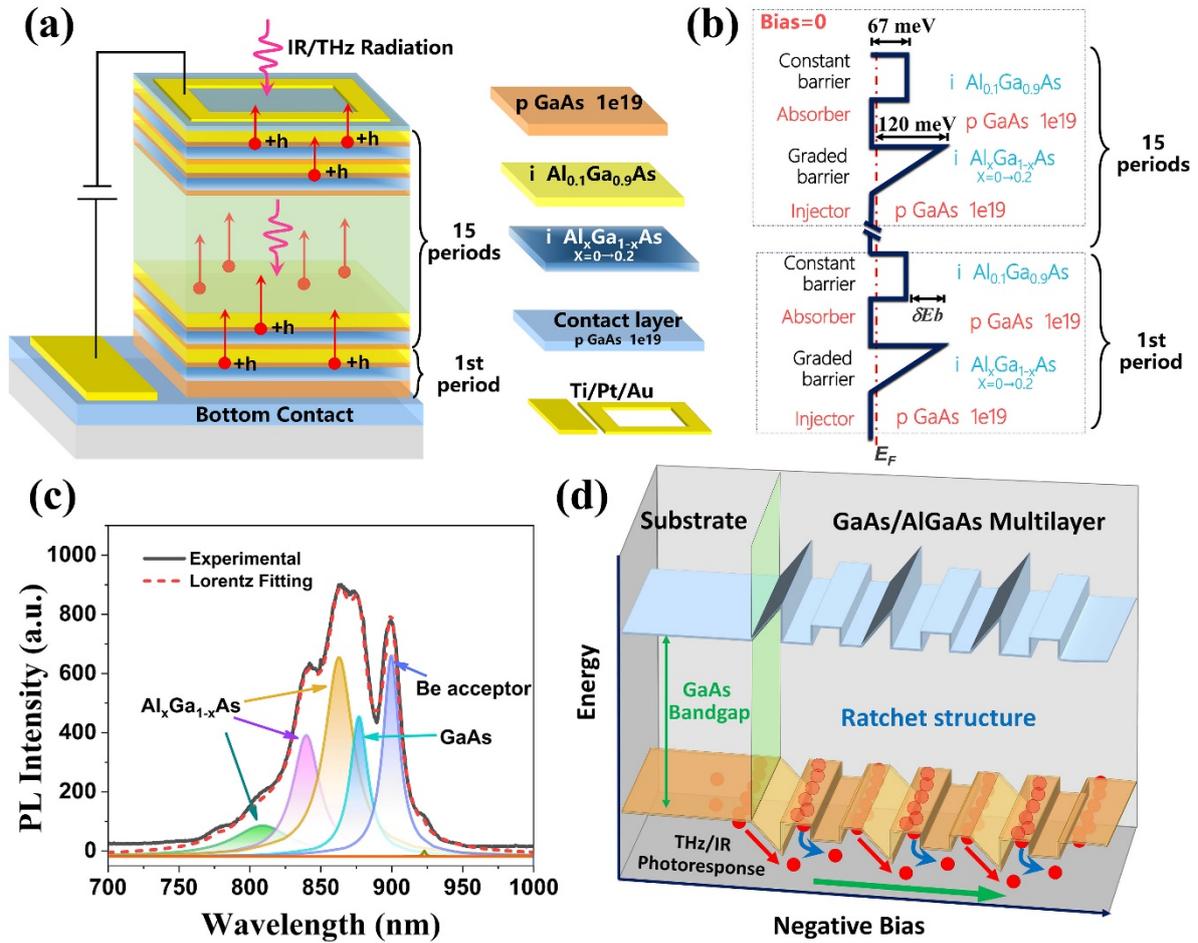

**Fig.1 (a). Structure of the ratchet photodetector device. (b). Band diagram of ratchet photodetector, the basic detection cell is displayed in the dashed box which is a four-layer structure consists of injector layer, graded barrier, absorber layer and constant layer. (c). The photoluminescence (PL) spectrum of the quantum ratchet photodetector at room temperature. (d). Schematic diagram of the THz/IR photoresponse in the ratchet structure under a negative bias, the red arrows indicate the electrical pumped hot holes injection and the blue arrows represent the THz absorption.**

The schematic working mechanism diagram of the THz/IR photoresponse in ratchet photodetector is shown in Fig.1(d). The IR or THz photon absorption is based on the ratchet-like valance band and occurs in the highly doped GaAs layers (injector and absorber) followed by internal photoemission. The emitted carriers then will be swept out and collected by the electrode. The red arrows in Fig.1(d) indicate the THz absorption related hot holes injection and the blue arrows represent the THz/IR absorption.

## 2.2 Dark current and ratchet effect

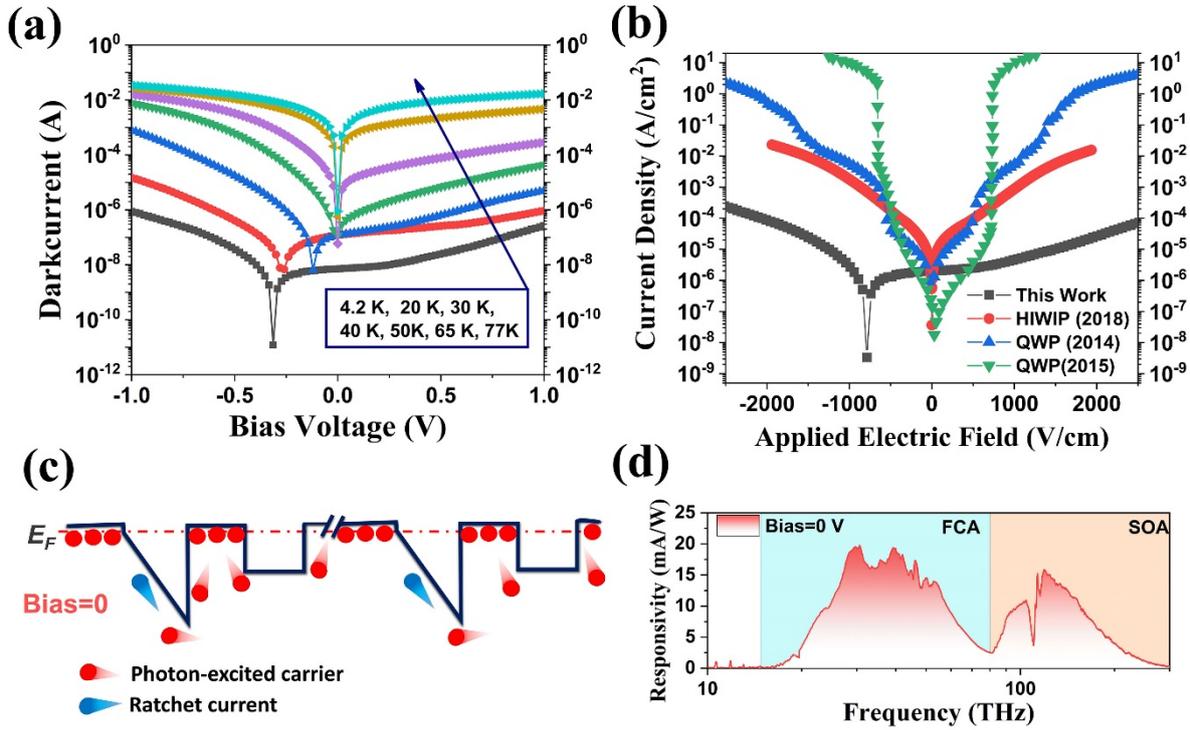

**Fig. 2. (a). The dark currents of the ratchet photodetector at different temperatures. (b). Dark currents of the ratchet photodetector (5K), QWP (2015, 7K) [13] QWP (2014, 8K) [35] and HIWIP (2018, 3.5K) [16] at the temperature around liquid helium temperature. (c). Valence-band diagram under zero bias: IR photons stochastically change the momentum and energy of holes both in injector and absorber first. Then the asymmetry of the repeat unit of the barrier potential bias the motion of holes to the left. (d). Responsivity of the ratchet detector in the whole responsive range with zero bias voltage at 4.2 K, the broadband response is due to the free carrier absorption (FCA) and split-off (SO) band absorption (SOA).**

The dark currents of the ratchet photodetector at different temperatures are shown in Fig.2(a). The I-V curves show rectification characteristics because of the asymmetric of the ratchet structure, which could be explained by the tilting ratchet effect [27]. Although there is a constant barrier between two ratchets, the applicability of tilting ratchet effect still holds. With a negative bias (with the top contact being grounded), the ratchet-like barrier will be tilted towards flat band so that the transportation of the holes is permitted. The permanent potential barrier of the graded barrier/absorber junction in the opposite direction could not be overcame under a positive bias, which caused the relatively low current compared with that of negative bias.

The dark currents of the ratchet photodetector, QWP [13, 35], and HIWIP [16] at the temperature around liquid helium temperature are presented in the Fig.2(b). It is clear that the dark current of the quantum ratchet photodetector is much lower than that of HIWIP/QWP over a large bias voltage range. We attribute this to the higher inherent barrier in the ratchet photodetector. The lowest barrier energy in the ratchet structure is about 67 meV (constant barrier), which is much higher than HIWIP and QWP. Most of the emitted and field emitted carriers are blocked by the barrier even with a large bias.

The ratchet-like potential barrier could not only result in low dark current, but also produce an evident ratchet current even under zero bias if the detector was illuminated by THz/IR radiation. Under zero bias (as is shown in Fig.2(c)), IR photons stochastically change the momentum and energy of holes both in injector and absorber first. The absorption occurred here include the free carrier absorption (FCA) and split-off band absorption (SOA) [36, 37]. Then the asymmetry of the repeated unit of the barrier potential bias the motion of holes to the left due to the asymmetric relaxation [25, 26]. This phenomenon is well known as the light-induced ratchet effect, which is regard as a novel photovoltaic mechanism [26]. Therefore, the ratchet photodetector can also be looked upon as a novel broadband photovoltaic detector. The responsivity of the detector with zero bias is displayed in Fig.2(d), which shows wide response range of 16-300 THz and performs a peak responsivity of 19.6 mA/W at 30 THz.

## 2.3 Bias-tunable THz/IR photoresponse

Fig.3(a) is the mapping result of the calibrated responsivity of the photodetector as a function with bias voltage (see Supplementary Information for the details of calibration method). The near infrared (NIR: 90-300 THz) photoresponse mechanism is shown in Fig.3(b), which is similar to the SO band photodetector and consists of three main steps: (I) photo-absorption exciting the holes from the highly doped emitters (injector and absorber), in which a direct transition occurs from L/H band to SO band, (II) scattering-assisted escape of the photo-excited carriers, and (III) collection of the escaped carriers [36] [37]. The deep valley around 110 THz in Fig.3(a) is from the hydroxyl absorption in the quartz window. The response mechanism of mid-infrared (MIR) and far-infrared (FIR) photons (15-90 THz) is mainly due to the FCA-based internal photoemission. In this range, injector behaves the same as the absorber and the detector works as a HEIWIP photodetector [34] (as shown in Fig.3(c)), in which the detector could response to the MIR or FIR photons and the cutoff frequency is typically dominated by the workfunction ($v_c=\Delta_B/h$). The permanent barrier of the absorber/graded barrier junction and the injector/costant barrier junction could not be overcame thus the cutoff frequency is difficult to reach below 15 THz. It should be note that the photoresponse mechanism is the same in the range of 15-300 THz under positive and negative bias, although the responsivity displays asymmetric behavior due to the asymmetric device structure.

Fig.3(a) exhibits completely different terahertz (4-20 THz) photoresponse spectrum shapes under positive and negative bias. The cutoff frequency of the photodetector under negative bias can reach as low as 4 THz, which breaks the limit of spectral response rule in semiconductor photoelectronic devices ($\Delta_B = 67$ meV). The bias dependent photocurrent produced by the 4-20 THz radiation from a 1000 K blackbody is shown in Fig.3(d). We find that the photocurrent under negative bias is higher than that of positive bias, which means more THz photons could be absorbed under negative bias. This result is in good agreement with the spectra in Fig.3(a). The transition bias in Fig.3(d) means that the ratchet current is counteracted at this value and then THz response will be enhanced under a larger negative bias. The deep valleys at around 8 THz and 11 THz in Fig.3(a) corresponds to the GaAs optical phonon and AlAs-like phonon absorption. The other small valleys at high frequencies are due to the multiple-phonons absorption. Thanks to the multi-period structure design, peak responsivity is as high as 50.3 mA/W, which is about 3 orders of magnitude higher than the OPHED.

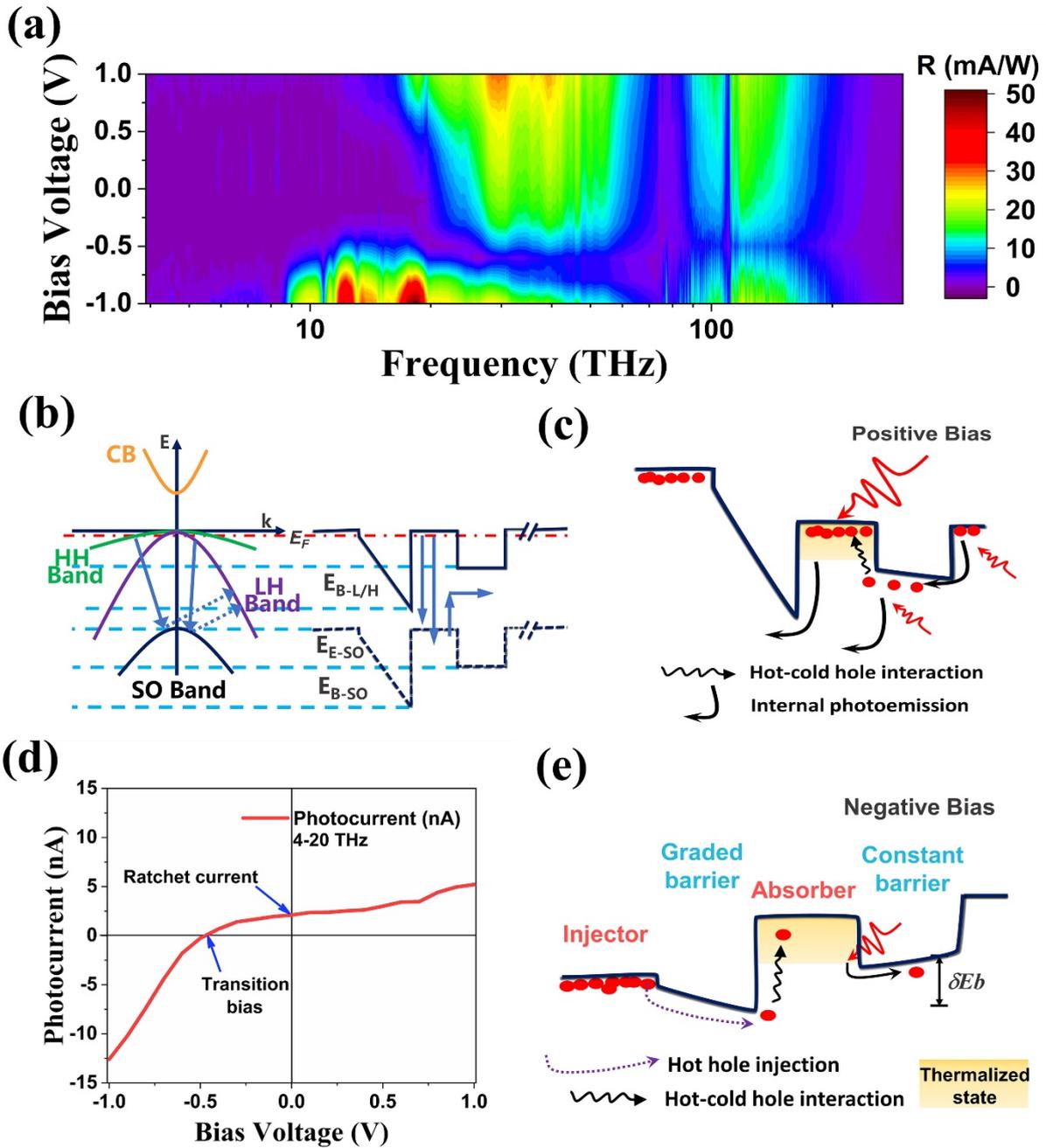

**Fig. 3. (a). The calibrated responsivity with different bias voltage on the whole responsive frequency range. (b). E-k diagram for highly doped emitter region (injector and absorber) of the photodetector and band diagram of the photodetector illustrating the NIR detection mechanism (SOA). (c). The schematic diagram of the MIR or FIR photons (with the energy higher than the internal barrier potential) response mechanism under positive bias. (d). Measured photocurrent produced by the 4-20 THz radiation from a 1000 K blackbody. (e). The schematic diagram of the THz photons (with the energy well below the internal barrier potential) response mechanism under negative bias.**

The bolometric effect or impurity-band absorption could not be the possible reason for the THz photon-response in the ratchet-like structure [20,21]. The total power density illuminated on the detector surface is estimated to be only about 50 μW/cm$^2$, which would not cause significant thermal effects. The temperature of the detector remains unchanged during measurement. In addition, the bolometric response is proportional to the temperature variation of the absorber upon photon absorption and the corresponding resistance changes monotonically with increasing bias. In contrast, the bias-dependent spectra of the ratchet detector exhibit completely different photoresponse behavior under negative and positive bias and the THz response only occurs at a large negative bias. So, the THz response is more likely to result from the electrical effect rather than thermal effect.

We think the THz response depends on the injected holes and occurs in the absorber. As is shown in Fig.3(e), the holes injected to the absorber are 'hot holes' because their higher energy compared to the indigenous cold holes or band edge of the absorber. Due to the energy difference ($\delta E_b$) between the two barriers, the hot-cold hole energy transfer will happen in the hot-hole relaxation process, which cause the redistribution of the holes in the absorber [20]. As a consequence, a high energy thermalized state will be formed and the holes then be able to response to the low frequency THz radiation. This mechanism is believed to be the main cause of the THz response. Different from the OPHED, there is no external optical excitation source needed in the ratchet photodetector.

The THz response related hot-holes mainly originate from the electrically pumping current, which is generally equal to the dark current. The ratchet-like graded barrier will be tilted towards a flat band (as shown in Fig. 3(e)) with increasing negative bias, which induces a higher pumping current. As a consequence, the hot carrier injection increases and the hot-cold hole interaction is also enhanced. More holes will occupy high energy state to form the thermalized state so that the cutoff frequency decreases and the THz photoresponse is enhanced (as presented in Fig.3(a)). Through hot-cold hole interaction, part of the pumping current helps the THz response other than contributing to dark current.

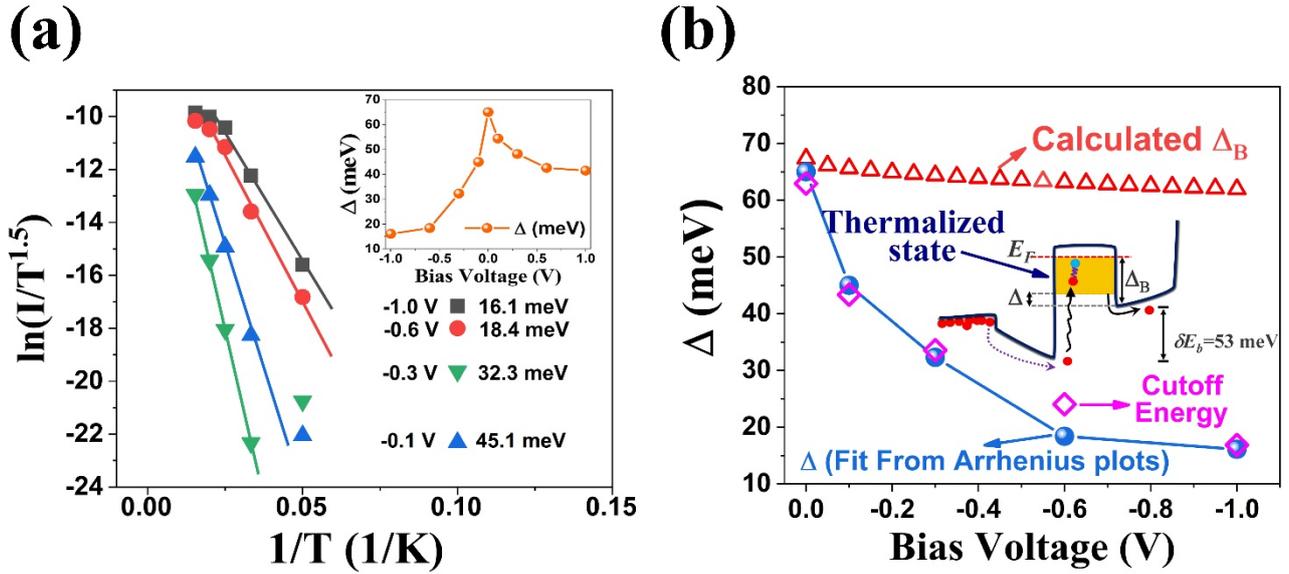

**Fig. 4. (a). Arrhenius plot for the ratchet photodetector at different bias voltage, and the bias dependent activation energy of the detector is presented in the inset. (b) Comparison between the activation energy (Δ) and the workfunction (Δ_B) of the absorber/constant barrier interface. The cutoff energies (defined as $h\nu_c$, $\nu_c$ is cutoff frequency) at different bias are also given. The inset in (b) indicate that the activation energy represents the energy difference between the thermalized state and band edge of the constant barrier.**

In order to further clarify the THz detection mechanism, we extract the activation energy of the detector (Δ) from the Arrhenius fitting based on the I-V-T data, as shown in Fig. 4(a). It can be seen that the activation energy agrees well with the cutoff energy (defined as $h\nu_c$, $\nu_c$ is cutoff frequency) given in Fig. 4(b), indicating that the THz response of the rachet detector is determined by Δ rather than the workfunction of the heterojunction interface (Δ_B) of the absorber/constant barrier interface. Fig. 4(b) shows the dependence of Δ, as well as Δ_B, on the bias voltage. It can be found that Δ is almost equal to Δ_B under zero bias. As the bias voltage increases from 0 to -1V, Δ_B is only reduced by 5 meV, while Δ is reduced greatly by 50 meV. At -1 V, Δ is only 16.1 meV, pushing the cutoff frequency to 4 THz and breaking the limit of Δ_B significantly. Such obvious difference comes from introduce of the graded rachet structure that leading to the electrically pumped hot hole injection and the energy level elevation of the thermalized state in the absorber as shown in the inset of Fig. 4(b). From this point of view, the extracted activation energy data exactly describes the energy difference between the thermalized state and band edge of the constant barrier. It should be noted that $\Delta_B$ is quite low and agree well with the $\Delta$ in previous studies of THz-HIWIP or THz-HEWIP, in which the corresponding mole fraction of Al in the GaAs/AlGaAs heterojunction is generally lower than 2% [16, 18, 34]. In contrast, a higher Al fraction could be used in ratchet photodetector.

The carrier-phonon coupling and carrier-carrier interaction play important roles in intervalence-band transition and formation of thermalized state [20, 33]. The electrically injected hot holes initially relax through carrier-phonon interaction followed by hole-hole scattering. Energy transfer between the injected hot holes and local cold holes causes a redistribution of energy for the holes in the absorber thereby forming a thermalized state. The lifetime of the thermalized state is about tens of picseconds scale [38, 39, 40]. Therefore, the thermalized holes could be able to be excited by absorbing THz photons.

In THz or infrared detection applications, a high value of responsivity is certainly desirable, but it is by no means the only practical consideration. Noise is also another key parameter to affect the performance of devices. It could thereby determine the noise equivalent power (NEP), the most widely quoted figure of merit for THz or infrared photodetectors. The noise theory of the photodetector could be found in the Supplementary Information. The calculated NEPs with the responsivities in Fig.3(a) for the whole response spectral range are shown in Fig.5(a). We can find that the NEPs for all frequencies are almost below the level of 100 pWHz$^{-1/2}$ and the optimal value is about 3.5 pWHz$^{-1/2}$ at 29 THz with a bias of -0.3 V. The reason why such a small NEP can be maintained at a large voltage is that the high potential barrier in the ratchet structure reduces the dark current of the device. In addition, part of the dark current will participate in the hot holes induced THz photoresponse and contribute to the THz photocurrent. The calculated results of $D_{PD}^*$ for the whole response spectral range are shown in Fig.5(b). We find that the $D^*_{PD}$ for all frequencies are almost above the level of $10^9$ cmHz$^{1/2}$/W and the optimal value is about $2.9\times10^{10}$ cmHz$^{1/2}$/W at 29 THz with a bias of -0.3 V.

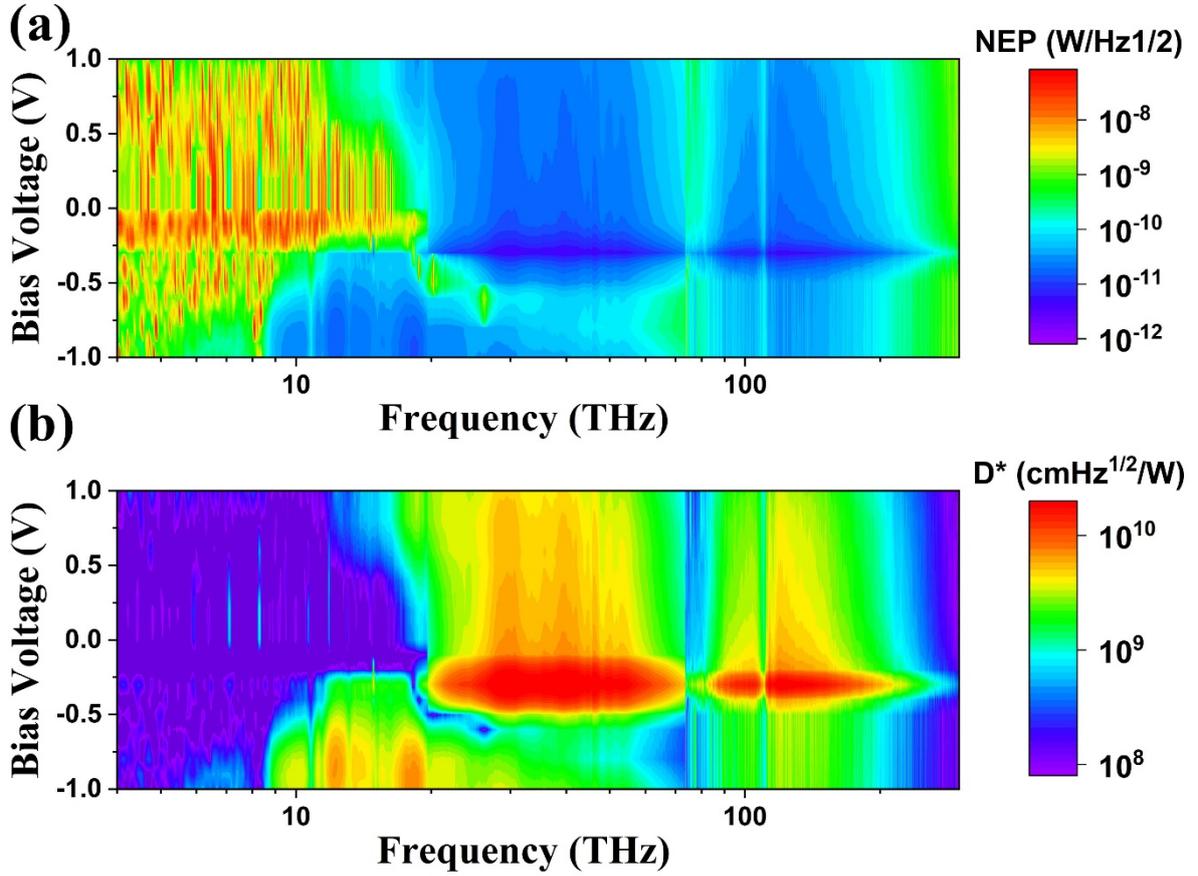

**Fig. 5. (a). The calculated noise equivalent power (NEP) under different bias voltage on the whole response spectral range (4-300 THz). (b). The calculated specific detectivity (D*) under different bias voltage on the whole response spectral range (4-300 THz).**

## 3. Discussion

In conclusion, we demonstrate an ultrabroadband THz/IR ratchet photodetector. The proposed ratchet photodetector presents potential superiorities compared with the existing photon-type THz photodetectors. The Table 1 shows a comparison of the ratchet detector with state-of-the-art GaAs-based detectors. The greatest advantage of the quantum ratchet photodetector is ultrabroadband photoresponse and it can allow normal incidence absorption thus bypassing the need for an extra optical coupler required for ISBT based detector (such as n-QWIP or QCD devices). The dark currents are significantly suppressed by the relatively high potential barrier in the ratchet structure，which enables the photodetector to work within a wide bias range and work at higher temperatures with further optimization. Inherent short carrier lifetime of the detector also makes it promising to operate with high speed. All features discussed above make the ratchet photodetector more promising in the application of broadband detection application and up-conversion pixelless imaging [17].

| Detector ID | Frequency Range | Peak Response | T(K) | Carrier Lifetime | D*(Jones) | Operation Mode | Ref. |
|---|---|---|---|---|---|---|---|
| QWP (mesa) | 5 THz | ~0.5 A/W | <15 K | <5 ps | $2.5 \times 10^{10}$ | **Polarization selection(45º)** electric drive | [13] |
| QWP (Patch antenna) | 5 THz | ~5.8 A/W | >5 K | <5 ps | $5 \times 10^{12}$ | normal incidence electric drive | [13] |
| QCD (mesa) | 3.5 THz | ~8.6 mA/W | 10 K | <5 ps | $5 \times 10^{7}$ | **Polarization selection(45º)** electric drive | [41] |
| QCD (Patch antenna) | 33 THz | ~50 mA/W | 83 K | <5 ps | $5.5 \times 10^{10}$ | normal incidence electric drive | [42] |
| HIWIP | 4-20 THz | ~6.8 A/W | 5 K | ---- | $2.3 \times 10^{10}$ | normal incidence electric drive | [17] |
| OPHED | 5-150 THz | ~69 µA/W | 5 K | ~150 fs | $1 \times 10^{9}$ | normal incidence electric drive & **optical pumping** | [20] |
| RP | 4-300 THz | ~50 mA/W | 5 K | ~150 fs | $2.9 \times 10^{10}$ | normal incidence electric drive | This work |

**Table 1. Comparison between the ratchet photodetector (RP) of the GaAs-based state-of-the-art detectors: quantum well infrared/THz detectors (QWIP)[13], quantum cascade detectors (QCD)[41,42], homojunction interface workfunction internal photoemission detector (HIWIP)[16], and optical pumped hot-hole effect detector (OPHED)[20].**

Besides, the photodetector also exhibits a bias-tunable behavior for THz photoresponse, which provides more solutions in some specific applications such as search-tracking, target recognition, THz/IR biophysics[43,44] and the research on the development of new THz/IR spectrometers. Furthermore, thanks to the multi-periods design in the ratchet structure, the responsivity of the quantum ratchet photodetector is about 3 orders of magnitude than that of OPHED. The electrically pumped THz photoresponse in the photodetector also enable it to abandon the requirement of the external excitation source in the OPHED that makes the photodetector much more compact. Most importantly, this work proves that the ratchet structure has a dramatic potential for the applications in terahertz or infrared photon detection. The photoresponse mechanism at zero bias in ratchet photodetector could be considered as a new photovoltaic scheme [25, 26], which also show a broadband photoresponse. This work not only demonstrates a novel broadband bias-tunable THz/IR photodetector, but also provides a new method to study the quantum ratchet or light-responsive ratchet.

## 4. Method

### 4.1 Wafer details and fabrication

The active region of the quantum ratchet photodetector consists of 16 periods basic ratchet detection cell directly grown by molecular beam epitaxy (MBE) on a 625 μm-thick semi-insulating GaAs substrate. The basic ratchet detection cell consists of a 20-nm-thick GaAs injector layer highly doped with Be to $1\times10^{19}$cm$^{-3}$, an 80-nm-thick Al$_x$Ga$_{1-x}$As graded barrier with x linearly varying from 0 (bottom) to 0.2 (top), a 20-nm-thick GaAs absorber layer highly doped with Be to $1\times10^{19}$cm$^{-3}$ and a constant Al$_x$Ga$_{1-x}$As barrier with the Al fraction of 0.1 and thickness of 80 nm. The active regions are sandwiched between two p-type GaAs ohmic contact layers doped with Be acceptor to $1\times10^{19}$cm$^{-3}$. The doping level of the injector and absorber is selected to realize high free carrier absorption while avoiding direct transitions from the heavy hole to light hole band. The high doping level also significantly increases the Fermi level in the device. The samples are processed using standard photo lithographic techniques. Square mesa structures with various areas of 400μm×400μm, 600μm×600μm, 1000μm×1000μm are fabricated using wet-chemical etching. The top electrical connection is a narrow ring contact (with a width of 70 μm) formed by deposition of Ti/Pt/Au using electron beam evaporation. The bottom common electrode is also p-contact metal made by Ti/Pt/Au.

## 4.2 Measurement details

The samples are mounted on 14 pin packages for electrical and optical measurements. All measurements were carried out at low temperatures. The photocurrent spectra at 4 K and different bias voltages shown in Fig. 3 are measured using a Fourier transform infrared spectrometer (Bruker VERTEX 80 IFS 66 v/s). The responsivities of the photodetector at different bias voltages are acquired using a calibrated blackbody (Infrared Systems Development Corporation IR-564/301) together with a low noise current preamplifier (Model SR570) and a lock-in amplifier (Model SR830). Mylar films beam splitter, KBr beam splitter and $GaF_2$ beam splitters are used to measure the frequency ranges of 4-20 THz, 10-150 THz and 50-300 THz, respectively. The window types of the optical cryostat for these three frequency ranges are HDPE window, KRS-5 window and Quartz window, respectively. The radiation beam illuminated on the device from the globar lamp in FTIR is a circular spot with 8 mm diameter. The aperture diameter of the blackbody is set as 0.2 inch, which is regarded as a point light source emitting divergent light illuminating on the device.

## Data availability

The data that support the findings of this study are available from the authors on reasonable request, see author contributions for specific data sets.

## Acknowledgements

This work was supported by the Natural Science Foundation of China (U1730246，12074249，11834011 and 62004209), Natural Science Foundation of Shanghai (19ZR1427000), Project funded by China Postdoctoral Science Foundation (2020M680458), Open Project funded by Key Laboratory of Artificial Structures and Quantum Control (2020-03). The authors would like to thank Prof. Yanfang Li and Dr. Zheng Shu for in-depth discussions.

## Author contributions

P. B. and W. C. conceived the experiment and designed the device structure. P. B. fabricated the photodetector devices and carried out the electrical and optical experiment. P. B., X. L., N. Y., and W. C. contributed to the data analysis and figures. X. L., X. B., S. H., Z. F., Z. T., H. L., J. C. contributed to the measurement of the quantum ratchet photodetector. L. L., E. H. L. grew the samples using molecular beam epitaxy. P. B. wrote the main manuscript text. X. L., W. C., Y. Z., N. Y., W. S., Y. X. and Z. Z. reviewed the manuscript. All authors discussed the results and contributed to the manuscript.

## Competing interests

The authors declare no competing financial interests.